\documentclass[twocolumn,prd,noshowpacs,nofootinbib,amsmath,amssymb,superscriptaddress,preprintnumbers]{revtex4-1}

\usepackage{graphicx,epsfig,psfrag,bm,amssymb}
\usepackage{mathrsfs,amsmath, amsfonts,hepunits, color}

\usepackage{dcolumn}
\usepackage{bm}
\usepackage{mciteplus}
\usepackage{tikz}
\usepackage{slashed}
\usepackage{slashed}

\definecolor{navy}{rgb}{0.2,0.2,0.7}
\usepackage[colorlinks]{hyperref}
\hypersetup{
     colorlinks  = true,
     citecolor  = navy,
 	linkcolor=red,
	urlcolor=navy
}

\newcommand{\be}{\begin{eqnarray}}
\newcommand{\ee}{\end{eqnarray}}
\def\lsim{\mathrel{\rlap{\lower4pt\hbox{\hskip 0.5 pt$\sim$}}
    \raise1pt\hbox{$<$}}}                
\def\gsim{\mathrel{\rlap{\lower4pt\hbox{\hskip1pt$\sim$}}
    \raise1pt\hbox{$>$}}}

\def\lsim{\mathrel{\rlap{\lower4pt\hbox{\hskip1pt$\sim$}}
    \raise1pt\hbox{$<$}}}
\def\gsim{\mathrel{\rlap{\lower4pt\hbox{\hskip1pt$\sim$}}
    \raise1pt\hbox{$>$}}}

\definecolor{Purple}{rgb}{0.8,0.2,0.5}

\begin{document}

\title{A New Flavor of Searches for Axion-Like Particles}
\author{Eder Izaguirre}
 \affiliation{Brookhaven National Laboratory, Upton, NY 11973}
\author{Tongyan Lin}
 \affiliation{Lawrence Berkeley National Laboratory, Berkeley, CA 94720}
  \affiliation{Berkeley Center for Theoretical Physics, University of California, Berkeley, CA 94720}
\author{Brian Shuve}
 \affiliation{SLAC National Accelerator Laboratory, Menlo Park, CA 94025}
 
 \preprint{SLAC-PUB-16876}

\begin{abstract}
We propose new searches for axion-like particles (ALPs) produced in flavor-changing neutral current (FCNC) processes. This proposal exploits the often-overlooked coupling of ALPs to $W^\pm$ bosons, leading to FCNC production of ALPs even in the absence of a direct coupling to fermions. Our proposed searches for resonant ALP production in decays such as $B\rightarrow K^{(*)}a,\,a\rightarrow \gamma\gamma$ and $K\rightarrow \pi a, \,a\rightarrow \gamma\gamma$ could greatly improve upon the current sensitivity to  ALP couplings to  Standard Model particles. We also determine analogous constraints and discovery prospects for invisibly decaying ALPs. 
\end{abstract}

\maketitle

{\noindent \it \bf Introduction:}~Axion-like particles (ALPs) are among the best-motivated candidates for particle extensions of the Standard Model (SM). ALPs arise as the Goldstone bosons of any theory with a Peccei-Quinn symmetry (PQ)  \cite{Peccei:1977hh,Weinberg:1977ma,Wilczek:1977pj}, which is a spontaneously broken global symmetry that is anomalous with respect to the SM gauge interactions. PQ symmetries, and hence ALPs, are ubiquitous in theories beyond the SM such as string theory \cite{Witten:1984dg,Conlon:2006tq,Svrcek:2006yi,Arvanitaki:2009fg} and supersymmetry \cite{Frere:1983ag,Nelson:1993nf,Bagger:1994hh}. ALPs were originally motivated by dynamical solutions to the strong-$CP$ problem \cite{Peccei:1977hh,Weinberg:1977ma,Wilczek:1977pj}, although  recent proposals suggest that ALP dynamics could also resolve the hierarchy problem \cite{Graham:2015cka}. 

Because ALPs are pseudo-Goldstone bosons, their properties are more constrained than arbitrary scalar fields. In particular, the shift invariance associated with the anomalous symmetry  protects  ALP masses from radiative corrections and so their masses can naturally take on any value. Furthermore, their interactions with SM fields arise  through higher-dimensional (or otherwise suppressed) operators, and these feeble couplings naturally situate ALPs within models of hidden sectors, either as the dark matter candidate itself \cite{Preskill:1982cy,Abbott:1982af,Dine:1982ah} or as a mediator through the ``axion portal'' \cite{Nomura:2008ru}. This has driven a comprehensive laboratory search program ranging from high-energy collider physics to condensed matter systems (see for example Refs.~\cite{Bergsma:1985qz,Riordan:1987aw,Bjorken:1988as,Blumlein:1990ay,Abbiendi:2002je,Davoudiasl:2005nh,Chou:2007zzc,Pugnat:2007nu,Afanasev:2008jt,Fouche:2008jk,Arik:2008mq,Asztalos:2009yp,Davoudiasl:2009fe,Ehret:2010mh,Jaeckel:2012yz,Budker:2013hfa,Aad:2014ioa,Mimasu:2014nea,Aad:2015bua,Jaeckel:2015jla,Dobrich:2015jyk,Knapen:2016moh,Marciano:2016yhf}). 

In the simplest models, ALPs couple to the SM gauge boson field strengths according to the anomaly of the corresponding PQ symmetry,
\be
\label{eq:lagrangian}
\mathcal{L} \supset - \frac{g_{aV}}{4}\,a\,V_{\mu\nu}\tilde{V}^{\mu\nu},
\ee
where $a$ is the ALP field, $\tilde{V}^{\mu\nu} = \epsilon^{\mu\nu\rho\sigma} V_{\rho\sigma}/2$ and $V_{\mu\nu}$ is the gauge boson field strength for a SM vector boson $V$. In principle, $a$ can interact with all three SM gauge fields, with the relative couplings determined by the gauge charges of the fermion fields giving rise to the PQ symmetry anomaly. Phenomenological studies of ALPs typically focus on their couplings to gluons and photons, since these largely determine the rates of ALP interactions at energies well below the electroweak scale. For example, the dominant constraints on ALPs for $M_a\sim\mathrm{MeV}-\mathrm{GeV}$ arise from beam-dump experiments  \cite{Bergsma:1985qz,Riordan:1987aw,Bjorken:1988as,Blumlein:1990ay,Dobrich:2015jyk} and high-energy colliders \cite{Abbiendi:2002je,Jaeckel:2012yz,Aad:2014ioa,Mimasu:2014nea,Aad:2015bua,Jaeckel:2015jla} through the coupling of ALPs to photons. By contrast, the couplings of ALPs to electroweak gauge bosons are often neglected because their effects are suppressed for energies $E\ll M_W$, and the corresponding constraints are presumed to be subdominant. 

\begin{figure}[t]
\centering
\includegraphics[width=0.3 \textwidth ]{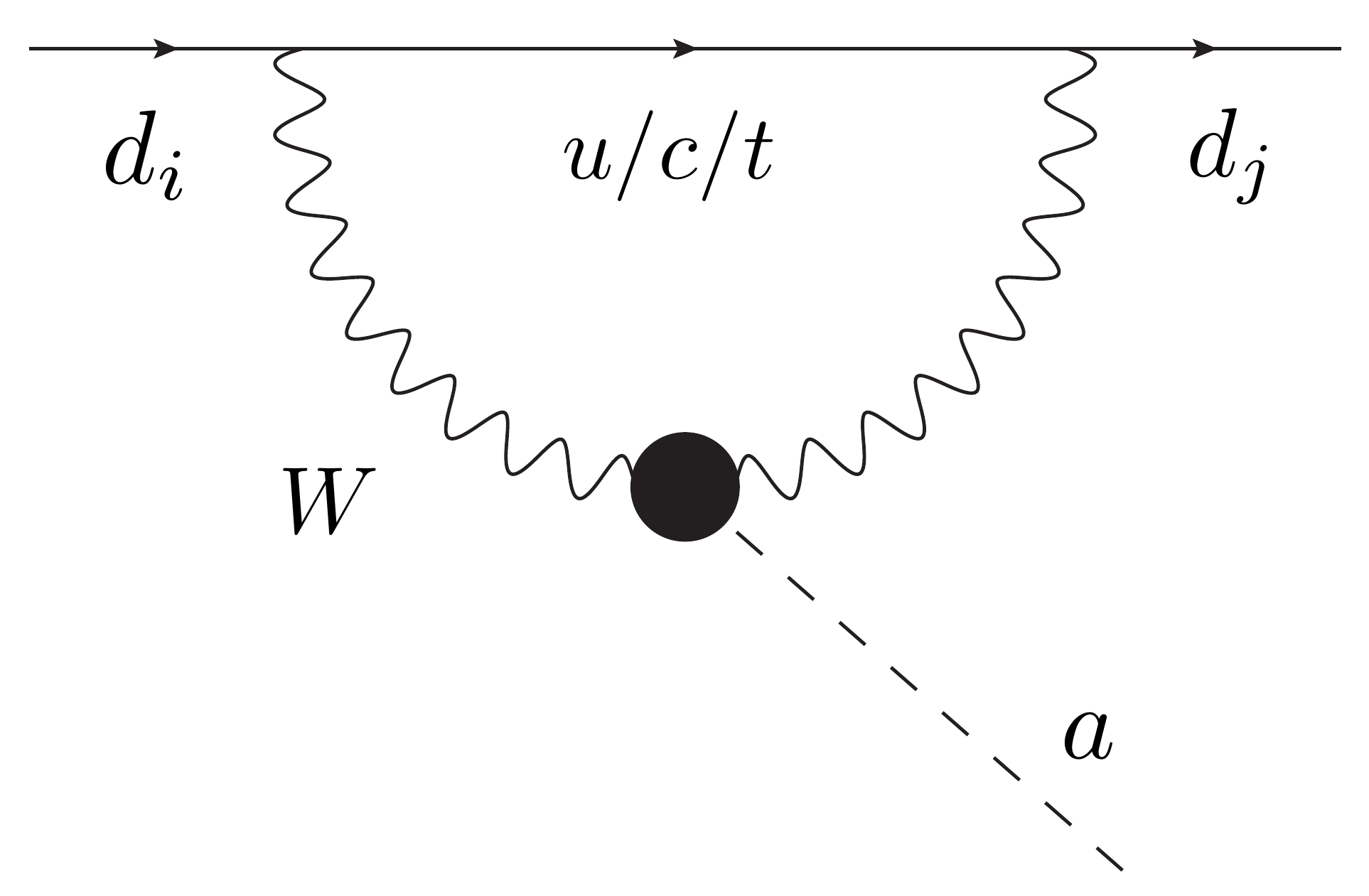}
\caption{Axion-like particle production in flavor-changing down-type quark decay, $d_i\rightarrow d_j+a$ .}
\label{fig:FCNC_feynman}
\end{figure}

Contrary to this lore, we show in this \emph{Letter} that couplings of ALPs to $W^\pm$ bosons can give rise to observable signatures and may, in fact, provide the best sensitivity to ALPs for masses below 5 GeV. ALPs with $aW\tilde{W}$ couplings can be emitted in flavor-changing neutral current (FCNC) processes such as those shown in Fig.~\ref{fig:FCNC_feynman}. In the scenario where the ALP couples predominantly to electroweak gauge bosons, the subsequent decay of the ALP into photons gives rise to signatures such as $B\rightarrow K^{(*)}a,\,a\rightarrow\gamma\gamma$. Since SM rates for such FCNC processes are small \cite{Glashow:1970gm}, ALP production in FCNC decays therefore provides a striking signature with excellent prospects for discovery.  Rare meson decays are already powerful probes of low-mass scalars possessing a direct coupling to quarks \cite{Wise:1980ux,Hall:1981bc,Frere:1981cc,Hiller:2004ii,Batell:2009jf,Freytsis:2009ct,Andreas:2010ms,Essig:2010gu,Dolan:2014ska,Krnjaic:2015mbs}, but our results show that ALPs can be discovered in these channels even if no direct coupling to quarks is present at leading order. Our proposed ALP signatures also predict the dominant $a\rightarrow\gamma\gamma$ decay instead of the fermionic decays which are most important when $a$ couples directly to SM fermions.\\

In the remainder of this {\it Letter}, we first present the ALP effective field theory (EFT). We subsequently derive the rates of ALP production in the most promising channels, namely $B\rightarrow K^{(*)}a,\,a\rightarrow \gamma\gamma$ and $K\rightarrow \pi a,\, a\rightarrow \gamma\gamma$. Next, we examine the prospects for direct ALP production at present and upcoming $B$-factories. Finally, we study the complementary scenario in which the ALP decays predominantly into  invisible states, determining the sensitivity of current and upcoming facilities to this possibility.\\

{\noindent \it \bf ALP Production in FCNC Decays:}~For concreteness, we consider a simplified model where the ALP  couples only to the field strengths of the $\mathrm{SU}(2)_W$ gauge bosons,
\be\label{eq:eft}
\mathcal{L}  = (\partial_\mu a)^2 - \frac{1}{2}M_a^2a^2 - \frac{g_{aW}}{4}\,a\,W^a_{\mu\nu}\tilde{W}^{a\mu\nu},
\ee
 where the $g_{aW}$ coupling is the leading term in the EFT expansion. This situation could arise if all fermions charged under the PQ symmetry possess only $\mathrm{SU}(2)_W$ gauge interactions, although models where $a$ additionally couples to the hypercharge gauge bosons give qualitatively similar results (see Appendix \ref{app:hypercharge}).  After electroweak symmetry breaking, the coupling $g_{aW}$ generates interactions between $a$ and $W^+W^-$, as well as $ZZ$, $Z\gamma$, and $\gamma\gamma$ in ratios given by the weak mixing angle.

We have computed the contribution of Eq.~(\ref{eq:eft}) to the amplitude for $d_i\rightarrow d_j a$ depicted in Fig.~\ref{fig:FCNC_feynman}. The result is replicated by the following effective interaction (assuming negligible up-quark mass):
\begin{align}
\mathcal{L}_{d_i\rightarrow d_j} &\supset - g_{ad_id_j}(\partial_\mu a)\,\bar d_j \gamma^\mu \mathcal{P}_{\rm L} d_i+\mathrm{h.c.},\label{eq:FCNC_coupling}\\
g_{ad_id_j} &\equiv -\frac{3\sqrt{2}G_{\rm F}M_W^2g_{aW}}{16\pi^2}\sum_{\alpha\in c,t}V_{\alpha i}V_{\alpha j}^*f(M_\alpha^2/M_W^2)\nonumber,\\
f(x) &\equiv \frac{x\left[1+x(\log x-1)\right]}{(1-x)^2}\nonumber,
\end{align}
where $G_{\rm F}$ is the Fermi constant and $V_{ij}$ are the relevant entries of the Cabibbo-Kobayashi-Maskawa (CKM) matrix. Note that $f(x)\approx x$ for $x\ll1$ such that the interaction is proportional to $M_\alpha^2/M_W^2$ for $M_\alpha\ll M_W$. There is an additional contribution to the effective coupling suppressed by factors of the external quark masses ($\sim M_{d_i}^2 / M_W^2$) that we have neglected to write in Eq.~(\ref{eq:FCNC_coupling}).

For flavor-changing couplings, the result is finite and depends only on the IR value of the effective coupling $g_{aW}$:~while individual diagrams in Fig.~\ref{fig:FCNC_feynman} are UV divergent, the divergences cancel when summed over intermediate up-type quark flavors. Because
the divergent terms are independent of quark mass, the unitarity of the CKM matrix requires that they sum to zero.
This is in contrast with models possessing a direct ALP-quark coupling, in which the FCNC rate is  sensitive to the  UV completion of the  theory \cite{Freytsis:2009ct,Batell:2009jf}.\\

{\noindent \it \bf Diphoton Searches for ALPs:}~We now discuss the prospects for the sensitivity of current and future probes to the ALP model in Eq.~(\ref{eq:eft}). We divide our discussion according to the two principal production modes:~secondary ALP production from rare decays of SM mesons, and primary ALP production at colliders.

ALP production in rare meson decays is, by far, the most promising new search mode. 
The quark coupling in Eq.~(\ref{eq:FCNC_coupling}) mediates FCNC decays of heavy-flavor mesons such as $B\rightarrow K^{(*)}a$ and $K \rightarrow \pi a$. To compute the rates of $B$-meson decays to pseudoscalar and vector mesons, we employ the hadronic matrix elements calculated using light-cone QCD sum rules \cite{Ball:2004ye,Ball:2004rg}. For $K^\pm\rightarrow\pi^\pm a$, we use the hadronic matrix element resulting from the Conserved Vector Current hypothesis   \cite{Gershtein:1955fb,Feynman:1958ty,Deshpande:2005mb} in the flavor-$\mathrm{SU}(3)$ limit assuming small momenta. The matrix element for $K^0\rightarrow \pi^0 a$ is related to that of $K^\pm\rightarrow \pi^\pm a$ by isospin symmetry, and so the matrix element for the $K_{\rm L}$ ($K_{\rm S}$) mass eigenstate is found by taking the imaginary (real) part of the $K^\pm\rightarrow\pi^\pm a$ matrix element \cite{Marciano:1996wy}. We keep only the leading terms from Eq.~(\ref{eq:FCNC_coupling}) that are unsuppressed by external momenta. The decay rates are:
\begin{align}
\Gamma(B\rightarrow Ka) & = \frac{M_B^3}{64\pi}\,|g_{abs}|^2\left(1-\frac{M_K^2}{M_B^2}\right)^2 f_0^2(M_A^2)\, \lambda_{Ka}^{1/2},\nonumber\\
\Gamma(B\rightarrow K^* a) &= \frac{M_B^3}{64\pi}\,|g_{abs}|^2\,A_0^2(M_a^2)\,\lambda^{3/2}_{K^* a},\nonumber\\
\Gamma(K^+\rightarrow \pi^+a) & = \frac{M_{K^+}^3}{64\pi}\left(1-\frac{M_{\pi^+}^2}{M_{K^+}^2}\right)^2 |g_{asd}|^2\,\lambda^{1/2}_{\pi^+ a}\nonumber,\\
\Gamma(K_{\rm L}\rightarrow \pi^0a) & = \frac{M_{K_{\rm L}}^3}{64\pi}\left(1-\frac{M_{\pi^0}^2}{M_{K_{\rm L}}^2}\right)^2 \,\mathrm{Im}(g_{asd})^2\,\lambda^{1/2}_{\pi^0 a},\nonumber
\end{align}
where $\lambda_{Ka} = \left[1-\frac{(M_a+M_K)^2}{M_B^2}\right]\left[1-\frac{(M_a-M_K)^2}{M_B^2}\right]$, along with analogously defined $\lambda_{K^* a}$, and $\lambda_{\pi^{+,0} a}$. $f_0(q)$ and $A_0(q)$ are appropriate form factors from the hadronic matrix elements, obtained from Refs.~\cite{Ball:2004ye} and \cite{Ball:2004rg}, respectively. 
For the $a$ mass range we study, $M_a\ll M_W$, the dominant decay mode is $a\rightarrow\gamma\gamma$. 

We begin our phenomenological study with the signature $B\rightarrow K^{(*)}a,\,a\rightarrow\gamma\gamma$, which has the best sensitivity to ALPs. While the same rare meson decay with $a\to \gamma \gamma$ is also predicted in models with pseudoscalars possessing only direct quark couplings \cite{Dolan:2014ska}, the diphoton mode is only dominant for ALP masses below the pion threshold in those scenarios. Moreover, to our knowledge, no such search has been carried out, nor has the SM continuum process $B\rightarrow K^{(*)}\gamma\gamma$ been previously measured \cite{Reina:1997my}. There are measurements of the processes $B\rightarrow K^{(*)}\pi^0,\,\pi^0\rightarrow\gamma\gamma$ at BaBar and Belle \cite{Abe:2006qx,Aubert:2007hh,BABAR:2011aaa,Duh:2012ie}, which are similar to our proposed ALP searches but are restricted to  $M_{\gamma\gamma}\sim M_{\pi^0}$. These branching ratios are measured with 2$\sigma$ uncertainties $\sim10^{-6}$, thus this value serves as a concrete benchmark for conservatively estimating the sensitivity to $B\rightarrow K^{(*)}a$. Since the ALP searches are a straightforward resonance search, however,  backgrounds can be estimated using sidebands, and we expect current and future $B$-factories will have even better sensitivity to $\mathrm{Br}(B\rightarrow K^{(*)}a)$.  

\begin{figure}[t]
\centering
\hspace{-0.5cm}
\includegraphics[width=0.5 \textwidth ]{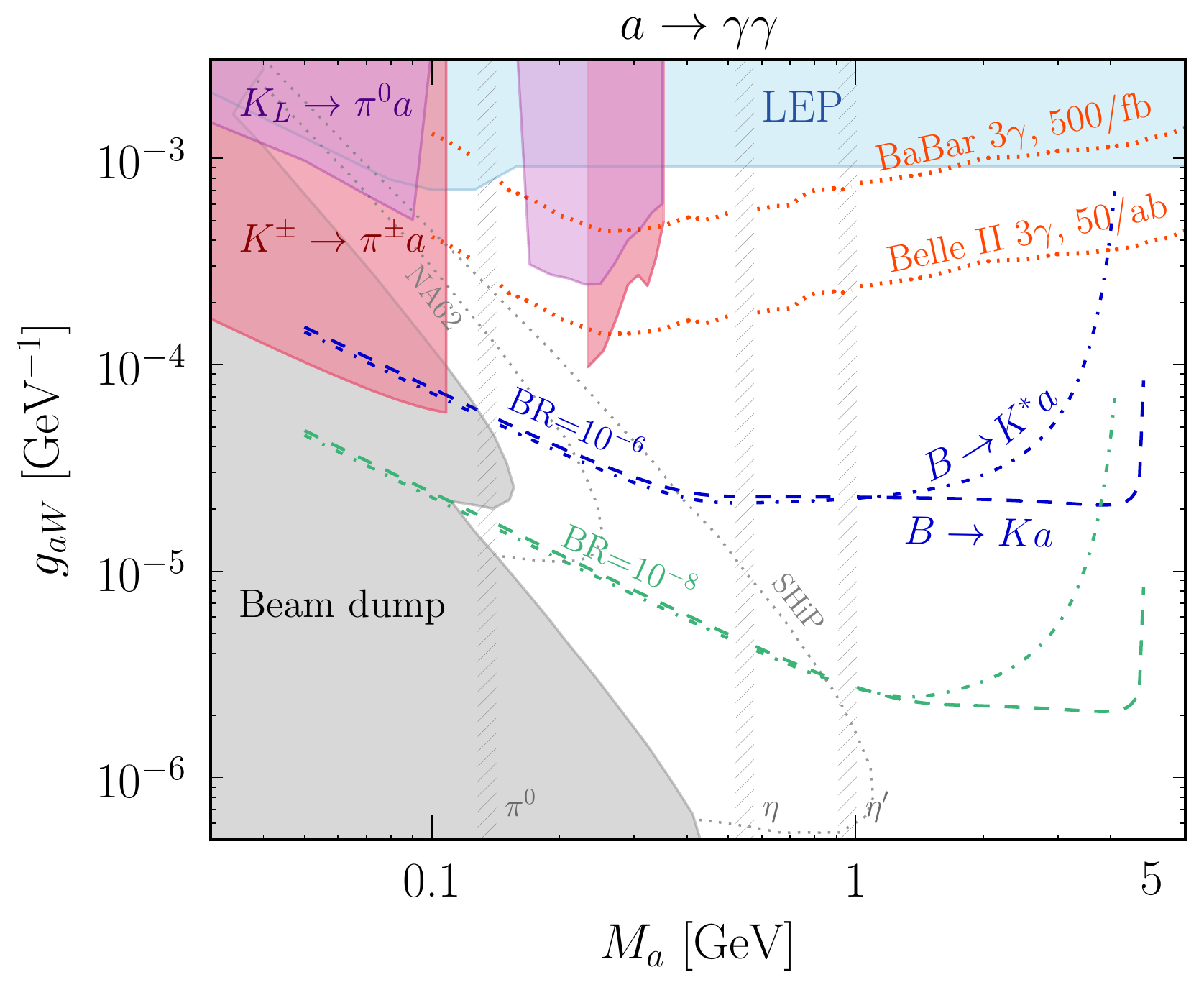}
\caption{Sensitivity of proposed searches in the ALP parameter space assuming the ALP decays primarily to two photons.   We show the reach of our proposed $B\rightarrow Ka$ (dashed lines)  and $B \rightarrow K^* a$ searches (dot-dashed lines), for sensitivity to branching ratios of $10^{-6}$ and $10^{-8}$. We also derive new constraints from rare $K$ decays from $K^+\rightarrow \pi^+ a$ (red) and $K_L \to \pi^0 a$ (purple). The dotted red lines indicate the reach we find from dedicated searches for $e^+ e^- \to a\gamma$ at current and upcoming $B$-factories. Existing projections are  shown with dotted lines for a proposed dedicated search in beam-dump mode at NA62 (gray), and for a recently-proposed beam-dump experiment, SHiP (gray). Shaded regions indicate current bounds from beam-dump experiments (gray) and LEP (light blue).}
\label{fig:BtoKa_reach}
\end{figure}

We therefore show in Fig.~\ref{fig:BtoKa_reach} our projections for ALP searches in $B^\pm\rightarrow {K^\pm}^{(*)}a,\, a\rightarrow \gamma \gamma$ for two branching fraction benchmark sensitivities ($10^{-6}$ and $10^{-8}$). We  do not consider ALP masses around the $\pi^0$, $\eta$, and $\eta'$ masses, and we conservatively require that the ALP decay  within $L < 30$ cm of the collision point to be observable. Current BaBar and Belle data, as well as the upcoming Belle II experiment \cite{Abe:2010gxa}, have the potential to improve sensitivity to the ALP coupling $g_{aW}$ by up to three orders of magnitude over current constraints, providing a clear motivation for new ALP searches in rare meson decays.

The ALP can also be produced in decays $K\rightarrow \pi a,\,a\rightarrow\gamma\gamma$, and we derive bounds that are indicated by the shaded red and purple regions in Fig.~\ref{fig:BtoKa_reach}. We extract the bounds from existing measurements of the  processes $K^\pm\rightarrow \pi^\pm \gamma\gamma$ and $K_{\rm L}\rightarrow \pi^0\gamma\gamma$ that have been carried out by E949 \cite{Artamonov:2005ru}, NA48/2, NA62 \cite{Ceccucci:2014oza}, and KTeV \cite{Abouzaid:2008xm}.  For $K^\pm\rightarrow\pi^\pm\gamma\gamma$, we obtain limits for $M_a > 100$ MeV using measurements by NA48/2+NA62 of the branching fraction  in bins of $M_{\gamma\gamma}$~\cite{Ceccucci:2014oza}, requiring that the signal cannot exceed the central value + $2\sigma$ in each bin. Taking into account the kaon beam energy, we further require that $a$ decay within 10 m of the $K^\pm$ decay vertex so that its photons are registered in the detector.  We use the E949 search for $K^\pm\rightarrow\pi^\pm \gamma\gamma$ \cite{Artamonov:2005ru} for $M_a \lesssim 100$ MeV, taking their bound on the partial branching fraction of $ 2.3 \times 10^{-8}$ for $p_\pi > 213$ MeV and requiring that $a$ decays within 80 cm of the stopped kaon. For the KTeV search in $K_{\rm L}\rightarrow\pi^0\gamma\gamma$ \cite{Abouzaid:2008xm}, we require that $a$ decay within 1~m of the primary $K_{\rm L}$ decay (given a detector resolution of $\approx0.3$ m \cite{Worcester:2009qt}), apply the provided signal acceptance and require that the ALP signal not exceed the observed number of events ($+2\sigma$) in each $M_{\gamma\gamma}$ bin. We emphasize that a dedicated sideband resonance search for ALPs in either channel could  improve the sensitivity to $K\rightarrow\pi a$ production. Neither search constrains ALP masses around $M_{\pi^0}$; while measurements of $K_{\rm L}\rightarrow\pi^0\pi^0$ at KTeV are, in principle, sensitive to ALPs around the $\pi^0$ mass \cite{Alexopoulos:2004sx}, they are subdominant to existing limits.

In addition to ALP production in meson decays, direct production of ALPs through their couplings to photons, at either lepton colliders or  proton beam-dump facilities,  is a promising possibility. At low-energy lepton colliders, the reaction $e^+ e^- \rightarrow \gamma a,\, a\rightarrow \gamma \gamma$ \cite{Mimasu:2014nea} can give a diphoton resonance in 3-photon final states. To our knowledge, such a search has not been carried out at $B$-factories. We compute the estimated sensitivity of dedicated searches at BaBar and Belle II to $g_{aW}$ in this final state (shown in Fig.~\ref{fig:BtoKa_reach}), accounting for the leading-order 3$\gamma$ background. The signal region consists of events with three photons ($E_\gamma > 200$ MeV, $-0.8< \cos{\theta_\gamma} <0.97$, and $\Delta R_{\gamma\gamma}>0.1$) and one photon pair with $M_{\gamma\gamma}$ within $\delta M_a$ of $m_a$. The mass resolution $\delta M_a$ varies from $7-70$ MeV at BaBar \cite{Ferber:2015jzj} and comparable resolution at Belle II  across the $100~\MeV < M_a < 10~\GeV$ range; these values are consistent with the BaBar $M_{\gamma\gamma}$ resolution at the $\pi^0$ mass rescaled to higher/lower masses \cite{Aubert:2009mc}. In addition to the $3\gamma$ search mode, we  also considered exclusive $e^+e^-\rightarrow e^+e^-a,\,a\rightarrow\gamma\gamma$ production \cite{Marciano:2016yhf} and found it to be subdominant to other channels. At proton fixed-target experiments, Ref.~\cite{Dobrich:2015jyk} proposed a dedicated run in beam-dump mode at NA62, as well as estimated the prospects for the recently proposed SHiP experiment \cite{Alekhin:2015byh}, and we show their projections for comparison in Fig.~\ref{fig:BtoKa_reach}.

 At high-energy lepton colliders, the ALP is highly boosted and photons from $a$ decay are merged, such that the signature $Z\rightarrow \gamma a, \,a\rightarrow\gamma\gamma$ is constrained by diphoton searches at LEP \cite{Jaeckel:2015jla}. LEP currently gives the strongest constraints on $a$ over much of the parameter space we consider, although it can easily be superseded by searches for $a$ in rare meson decays.\\

\begin{figure}[t]
\centering
\hspace{-0.5cm}
\includegraphics[width=0.5 \textwidth ]{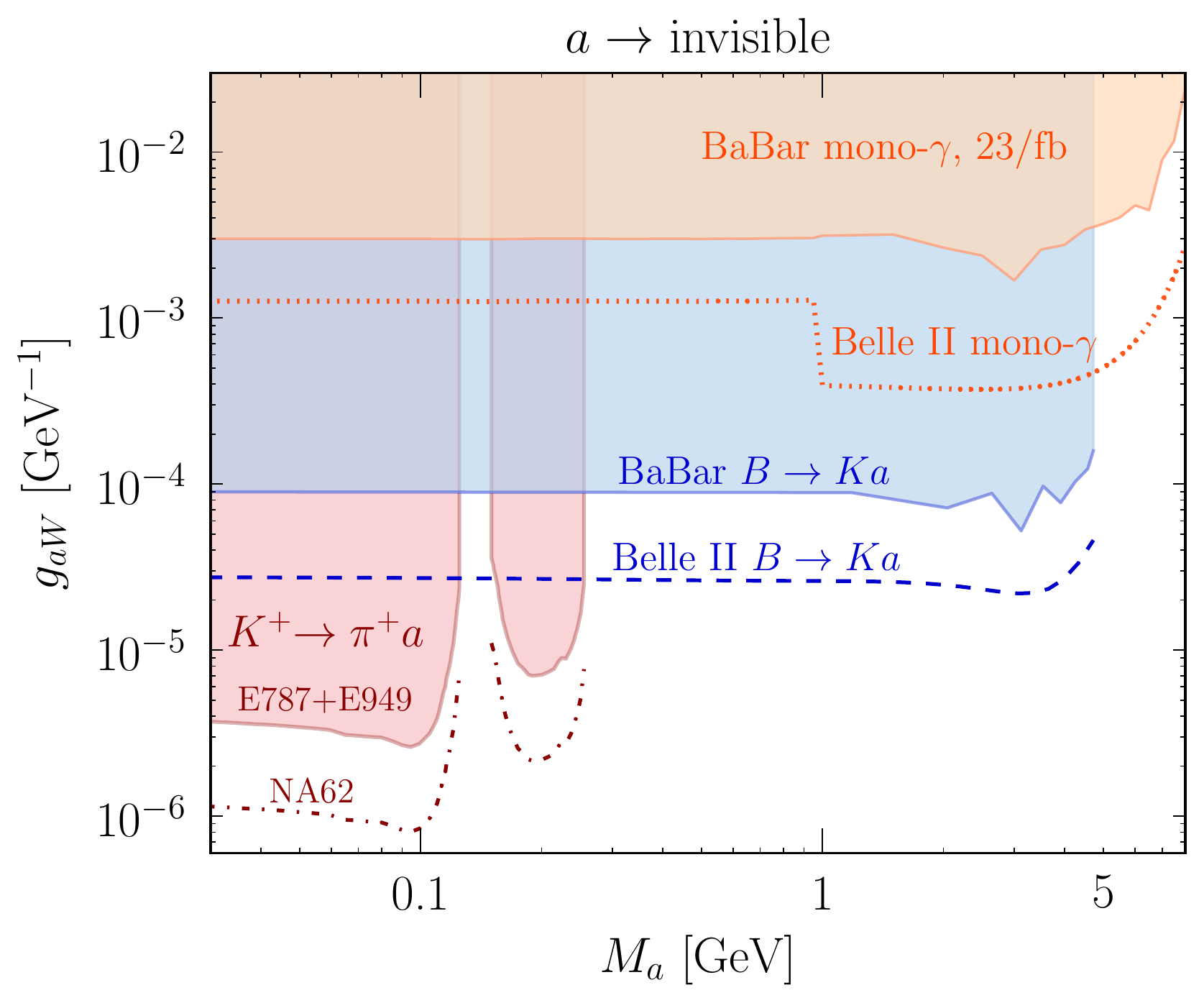}
\caption{Sensitivity of existing and planned searches to the ALP parameter space assuming the ALP decays invisibly.  We apply a BaBar search for $B \to K \nu \bar \nu$ to constrain the decay $B \to K a$ (shaded blue); this bound can be further improved with Belle~II (dashed blue). Similarly, data from E787 and E949 is used to constrain $K \to \pi a$ in two mass ranges (shaded red), with expected improvements from NA62 (dot-dashed red). We show bounds on $e^+ e^- \to a \gamma$ from a BaBar mono$-\gamma$ search (shaded orange) and the estimated reach for the same search at Belle II (dotted).
}
\label{fig:invisibleALP}
\end{figure}

{\noindent \it \bf The Invisible ALP:}~Up to this point, we have assumed that the ALP is produced and decays through the minimal interaction given in Eq.~(\ref{eq:eft}). However, since ALPs are relatively weakly coupled to SM particles, they are also excellent candidates for mediators between the SM and hidden sectors. If the hidden-sector particles are lighter than $a$, the ALP can have a large branching fraction to invisible states. Our results for invisibly decaying ALPs are summarized in Fig.~\ref{fig:invisibleALP}.

Invisible ALP  production in rare meson decays can be detected via missing mass and/or momentum. A promising mode is the $B\rightarrow K + \rm{invisible}$ reaction~\cite{Bird:2004ts,Kamenik:2011vy}.  A recent BaBar search for  $B \rightarrow K \nu \bar \nu$
reported sensitivity to this final state at the level of $10^{-5}$ in branching fraction \cite{Lees:2013kla}, and provided limits 
in bins of $M_{\rm inv}^2/M_B^2$ with a maximum value of $M_{\rm inv}^2/M_B^2 \leq 0.8$. 
We applied the results of the seach for the $B^+\rightarrow K^+ \nu \bar \nu$ final state to $B^+\rightarrow K^+ a$, taking the $2\sigma$ upper limit from the appropriate bin for a given $M_a$. This result is shown in Fig.~\ref{fig:invisibleALP}, along with a projection for Belle II, where we assume that the statistical uncertainty dominates in the measurement.

We also derive a 90$\%$ CL limit on ALP production in  $K\rightarrow \pi a,\, a \rightarrow \rm{invisible}$ (shown in Fig.~\ref{fig:invisibleALP}) using the results of E787 and E949 ~\cite{Artamonov:2009sz}, which searched for the SM process $K^\pm\rightarrow \pi^\pm \bar\nu \nu$ in two separate momentum bins \cite{Anisimovsky:2004hr,Artamonov:2008qb}. We include a projection for the results of the upcoming NA62 experiment, the goal of which is to observe 80 signal events for $K^\pm\rightarrow\pi^\pm\bar\nu\nu$ with very high signal purity \cite{NA62}.  In our projection, we scale the E787/E949 results by the ratio of the uncertainty on the SM process, assuming only statistical uncertanties for NA62.

Direct production of invisibly-decaying ALPs at lepton colliders arises from processes such as $e^+e^-\rightarrow \gamma a,\,a\rightarrow\mathrm{invisible}$ via the ALP couplings to $\gamma \gamma$ and $\gamma Z$. An existing monophoton and missing momentum search at BaBar constrains invisible ALPs within kinematic reach. We re-interpret the results of a search for untagged $\Upsilon(3S)\rightarrow \gamma A^0, \,A^0\rightarrow \rm{invisibles}$  from Ref.~\cite{Aubert:2008as} (for more details, see Refs.~\cite{Izaguirre:2013uxa,Essig:2013vha}). 
We find a limit from BaBar of $g_{aW}\sim(\mathrm{500~\GeV})^{-1}$, as shown in Fig.~\ref{fig:invisibleALP}. Moreover, we estimate that Belle II will extend coverage to $g_{aW}\sim(\mathrm{2~\TeV})^{-1}$ for $M_a > 1 $ GeV, where the search is statistics-limited. For $M_a < 1$ GeV, there is a large systematic error and the improvement in sensitivity for Belle II is less pronounced, although this could be ameliorated by improvements in background estimation methods. We find that production of ALPs in meson decays provide superior sensitivity for $M_a < M_b$, while monophoton searches provide complementary sensitivity above the $B$ mass.  Analogously, LEP monophoton and missing momentum searches provide complementary coverage at still larger masses \cite{Fox:2011fx}, although we find these searches are subdominant to $B$-factories for ALP masses below   $\sim10$ GeV.\\

{\noindent \it \bf Conclusions and Discussion:} In this {\it Letter}, we have studied the overlooked coupling of axion-like particles to $W^\pm$ bosons. We find that ALPs in the $10~\MeV < M_a \lsim 10~\GeV$ mass range can be exquisitely probed with current and upcoming low-energy, high-intensity accelerator experiments. In particular, rare FCNC meson decays, along with dedicated direct searches for ALP production at $B$-factories, have the potential to improve sensitivity to ALP-SM couplings by almost three orders of magnitude.

We have restricted our study to the effective interaction shown in Eq.~(\ref{eq:FCNC_coupling}), which is independent of the specific UV completion of the EFT in Eq.~(\ref{eq:eft}). However, we note that additional direct couplings of the ALP to SM fermions can be generated by renormalization-group evolution from the UV cutoff, resulting in cutoff-dependent contributions to $g_{ad_id_j}$   \cite{Freytsis:2009ct,Batell:2009jf}. The cutoff $\Lambda$ satisfies $\Lambda\sim \alpha_W g_{aW}^{-1}$, and we find that the UV-dependent contributions to ALP production are always subdominant to the UV-independent coupling in Eq.~(\ref{eq:FCNC_coupling}) for our  parameter space. These UV-dependent couplings could, however,  induce very rare ALP decays such as $a\rightarrow\mu^+\mu^-$, which could be discovered in future $B\rightarrow K^*a,\,a\rightarrow\mu^+\mu^-$ searches and  allow for a determination of the UV scale in combination with measurements in the diphoton channel.

Finally, the portal studied in this {\it Letter} is ripe for exploration at high-energy hadron colliders due to the enhanced coupling of the ALP to electroweak gauge bosons and rates that grow with energy in the EFT. Since high-energy probes can depend on  the UV completion of the  theory, it is beyond the scope of the low-energy probes proposed here and we  leave them for future study \cite{futurework}.

{\em Acknowledgments.} We are grateful to Daniele Alves, Nikita Blinov, Bertrand Echenard, Chris Hearty, Valentin Hirschi, Zoltan Ligeti, William Marciano, Amarjit Soni, and Jesse Thaler  for helpful conversations.  EI is supported by the United States Department of Energy
under Grant Contract de-sc0012704. TL is supported by the DOE under contract DE-AC02-05CH11231 and by NSF grant PHY-1316783. BS is supported by the DOE  under contract DE-AC02-76SF00515. Digital data related to our results can be found at \url{https://quark.phy.bnl.gov/Digital_Data_Archive/ALPSFlavor112016}.

\appendix
\section{Constraints on ALPs with hypercharge coupling}\label{app:hypercharge}

\begin{figure*}[t]
\centering
\hspace{-0.5cm}
\includegraphics[width=0.5 \textwidth ]{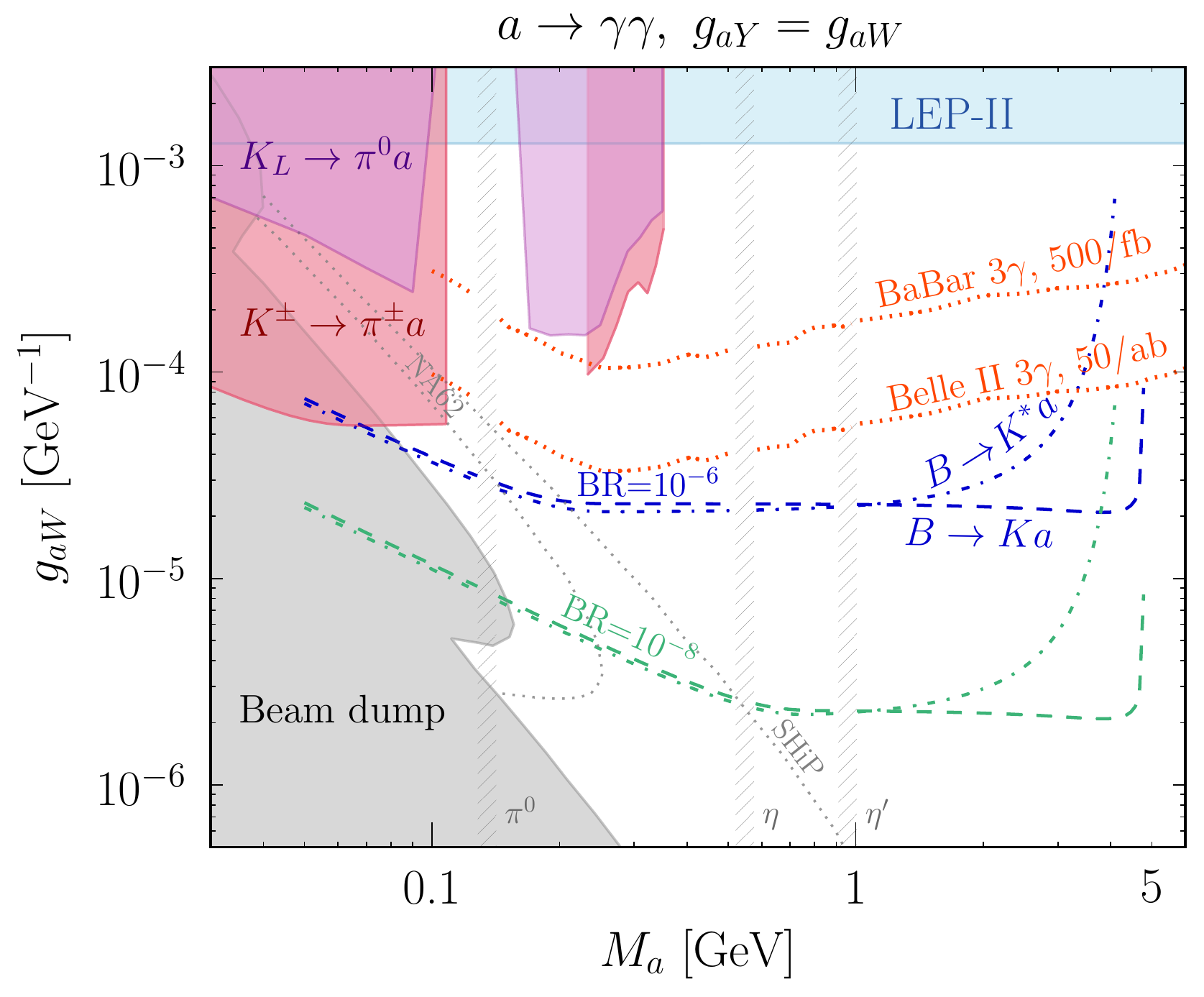}
\includegraphics[width=0.5 \textwidth ]{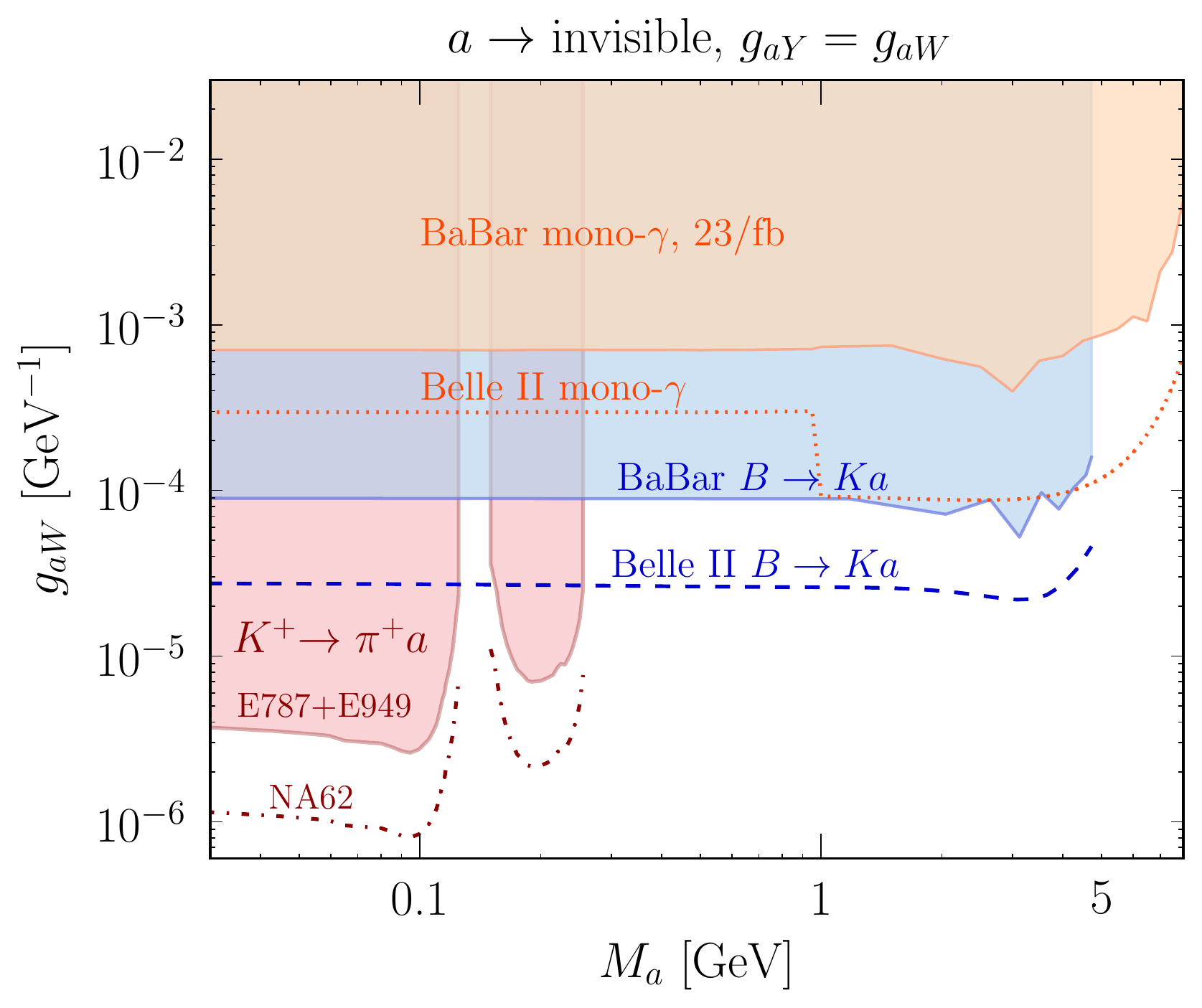}
\caption{Similar to Figs.~\ref{fig:BtoKa_reach}-\ref{fig:invisibleALP}, except we include a coupling of the ALP to hypercharge gauge bosons and set $g_{aY} = g_{aW}$.
}
\label{fig:ALPgY}
\end{figure*}

 In Fig.~\ref{fig:ALPgY}, we show the analogous results for ALPs that additionally couple to hypercharge gauge bosons. We take $g_{aY} = g_{aW}$ as a benchmark, such that the coupling of the ALP to photons is $g_{aW}$.   For $a \to \gamma \gamma$, the reach for rare meson decays is strengthened at low $M_a$, where the larger ALP decay rate leads to a larger fraction of decays within the detector volume.  The LEP constraint on $Z \to a \gamma$, shown in Fig.~2, is significantly weaker for $g_{aY} = g_{aW}$. Instead, we show the stronger LEP-II constraint on $e^+ e^- \to a \gamma$ taken from Ref.~\cite{Knapen:2016moh}. Even with the additional ALP coupling to hypercharge, however, we see that the projections and limits from rare meson decays can exceed the sensitivity of projected searches based on photon couplings alone.

\bibliography{ALP_FCNC}

\end{document}